\documentclass[conference]{IEEEtran}
\IEEEoverridecommandlockouts
\usepackage{amsmath,amssymb,amsfonts}
\usepackage{placeins}
\usepackage[utf8]{inputenc}
\usepackage{graphicx}
\usepackage{epstopdf}
\usepackage{booktabs}
\usepackage{textcomp}
\usepackage{bm}
\usepackage{tcolorbox}
\usepackage{braket}
\usepackage{amsthm}
\usepackage[textsize=tiny]{todonotes}
\usepackage{yquant}
\usepackage{hyperref}
\usepackage{mathtools}
\usepackage{circuitikz}
\usepackage{tikz}
\usetikzlibrary{angles, arrows.meta}
\usepackage{tkz-euclide}
\usepackage{float}
\usepackage{caption}
\usepackage{subcaption}
\usepackage{tabularx}
\usepackage{csvsimple}
\usepackage{xcolor}
\usepackage{comment}
\usepackage{gensymb}
\usepackage{multirow}
\usepackage{svg}
\usepackage{algorithm} 
\usepackage{algpseudocode} 
\usepackage{csquotes}
\usepackage{array}
\usepackage{adjustbox}

\usetikzlibrary{shapes.geometric, positioning, quantikz}

\usepackage[font=footnotesize]{subcaption}
\usepackage{yquant}
\useyquantlanguage{groups}
\usepackage{pgfplots}

\newtheorem{example}{Example}

\usepackage[style=ieee, maxnames=4, minnames=1, date=year, doi=false,isbn=false,backend=biber]{biblatex}

\usetikzlibrary{calc}

\def\BibTeX{{\rm B\kern-.05em{\sc i\kern-.025em b}\kern-.08em
    T\kern-.1667em\lower.7ex\hbox{E}\kern-.125emX}}

\DeclareFieldFormat
  [misc, book]
  {title}{\mkbibquote{#1}}

\AtEveryBibitem{
  \clearname{editor}%
  \clearfield{series}%
  \clearfield{isbn}%
  \clearfield{issn}%
  \clearfield{volume}%
  \clearfield{number}%
  \clearfield{pages}%
  \clearfield{url}%
  \clearfield{doi}%
}

\def\authorwidth{0.7cm}
\begin{document}

\title{Improving Hardware Requirements\\for Fault-Tolerant Quantum Computing\\by Optimizing Error Budget Distributions}

\author{
	\IEEEauthorblockN{ Tobias V. Forster\IEEEauthorrefmark{1}\hspace*{\authorwidth}
Nils Quetschlich\IEEEauthorrefmark{1}\hspace*{\authorwidth}
Mathias Soeken\IEEEauthorrefmark{2}\hspace*{\authorwidth}
Robert Wille\IEEEauthorrefmark{1}\IEEEauthorrefmark{3}\IEEEauthorrefmark{4}
}
\IEEEauthorblockA{\IEEEauthorrefmark{1}Chair for Design Automation, Technical University of Munich, Germany}
\IEEEauthorblockA{\IEEEauthorrefmark{2}Microsoft Quantum, Switzerland}
\IEEEauthorblockA{\IEEEauthorrefmark{3}Munich Quantum Software Company GmbH, Garching near Munich, Germany}
\IEEEauthorblockA{\IEEEauthorrefmark{4}Software Competence Center Hagenberg GmbH (SCCH), Austria}
\IEEEauthorblockA{\{t.forster, nils.quetschlich, robert.wille\}@tum.de\hspace*{\authorwidth}mathias.soeken@microsoft.com
\\
\url{https://www.cda.cit.tum.de/research/quantum}\hspace*{12mm}\url{https://quantum.microsoft.com}}
}

\maketitle

\begin{abstract}

Despite significant progress in quantum computing in recent years, executing quantum circuits for practical problems remains challenging due to error-prone quantum hardware. 
Hence, quantum error correction becomes essential but induces significant overheads in qubits and execution time, often by orders of magnitude.
Obviously, these overheads must be reduced. 
Since many quantum applications can tolerate some noise, end users can provide a maximum tolerated error, the error budget, to be considered during compilation and execution.
This error budget, or, more precisely, its distribution, can be a key factor in achieving the overhead reduction.
Conceptually, an error-corrected quantum circuit can be divided into different parts that each have a specific purpose. 
Errors can happen in any of these parts and their errors sum up to the mentioned error budget---but how to distribute it among them actually constitutes a degree of freedom.
This work is based on the idea that some of the circuit parts can compensate for errors more efficiently than others. 
Consequently, these parts should contribute more to satisfy the total error budget than the parts where it is more costly. 
However, this poses the challenge of finding optimal distributions. 
We address this challenge not only by providing general guidelines on distributing the error budget, but also a method that automatically determines resource-efficient distributions for arbitrary circuits by training a machine learning model on an accumulated dataset.
The approach is evaluated by analyzing the machine learning model's predictions on so far unseen data, reducing the estimated \mbox{space-time} costs for more than 75\% of the considered quantum circuits, with an average reduction of 15.6\%, including cases without improvement, and a maximum reduction of 77.7\%.

\end{abstract}

\vspace{-1mm}
\section{Introduction}

Quantum computing~\cite{nielsen2010quantum} has advanced substantially in both hardware and software over the last years.
This increasingly sparks interest in developing quantum applications in both academia and industry across different areas like \mbox{health-care}~\cite{santagati2024drug, fi15030094}, chemistry~\cite{motta2022emerging, cao2019quantum}, and others~\cite{herman2023quantum, orus2019quantum, 10821384}.
To realize such an application, the respective problem is encoded into a quantum circuit that, when executed, provides the solution to the problem.
However, the error-prone quantum hardware causes errors in the results, becoming more severe with increasing circuit sizes and leading, in the worst case, to unusable results for practically relevant problems. 

Thus, various quantum error correction techniques (such as, \mbox{e.g., \cite{gottesman1997stabilizer, Kitaev_1997, steane1999efficient}}) as well as corresponding methods and tools (such as, \mbox{e.g., \cite{PRXQuantum, berent}}) exist to counteract this behavior.
However, these techniques induce drastic overheads in the physical resource requirements that often increase, e.g., the number of physical qubits by orders of magnitude.
Obviously, these overheads need to be reduced significantly.

In fact, many quantum computing applications can tolerate a certain degree of error in the results.
To this end, the end user defines the maximum tolerated errors, the error budget, that can be considered during the compilation and execution of a \mbox{fault-tolerant} quantum circuit.
Moreover, since a \mbox{fault-tolerant} circuit consists of different parts, the error budget can be distributed among these parts.
Since distributing the budget manually is challenging, a uniform distribution of this over these parts is assumed.
However, allowing for more \mbox{errors in} one part and compensating for this with fewer errors in another part can save resources while keeping the overall error budget constant.
This constitutes a degree of freedom to distribute the error budget so that fewer hardware resources are needed, while, at the same time, this presents the challenge of how to find such distribution.
Hence, a better understanding of the influences of different error budget distributions on the resource requirements and a method to automatically determine good distributions for a given quantum circuit are required.

In this work, we address both of these challenges by proposing an approach that utilizes a supervised machine learning model to predict optimized error budget distributions for a given quantum circuit. 
To this end, a labeled dataset is accumulated by randomly sampling different error budget distributions, estimating their resource requirements for various quantum circuits, and saving the best distribution for each quantum circuit.
Then, the machine learning model is trained to learn the relationship between quantum circuits and corresponding optimized error budget distributions.

The evaluation of this approach provides general \mbox{guidelines---obtained} from the accumulated \mbox{dataset---on how} to distribute the error budget in a resource-efficient way.
Furthermore, the predicted error budget distributions by the machine learning model are compared to the usually assumed uniform distribution.
Here, the estimated \mbox{space-time} costs could be reduced for over $75 \%$ of the 383 considered quantum circuits with an average reduction of $15.6 \%$ (including samples without improvement) and a maximum reduction of $77.7 \%$.

\vspace{10mm}

The remainder of this work is structured as follows. 
\autoref{sec:Background} reviews the basics of quantum computing and \mbox{fault-tolerant} quantum computing with a subsequent review of resource estimation.
\autoref{sec:Motivation} motivates the proposed approach before its implementation is described in \autoref{sec:Optimizing the Error Budget Distribution}.
A comprehensive investigation of the approach and its results are provided by \autoref{sec:Evaluation}.
Finally, \autoref{sec:Conclusion} concludes this work.

\vspace{-5mm}
\section{Background}
\label{sec:Background}

This section briefly describes concepts relevant to understanding the principles of this work.
First, a general review of quantum computing is provided before we cover the basic concepts of fault-tolerant quantum computing and resource estimation for fault-tolerant quantum computing.

\subsection{Quantum Computing}
\label{subsec:Quantum Computing}

Quantum computing is an emerging computing technology that, instead of classical bits, uses quantum bits, or \emph{qubits}, that extend the capabilities of classical bits. 
In contrast to their classical counterparts and by exploiting the quantum nature of qubits, their states can also be a \emph{superposition} of 0 and 1.
The state of a qubit can be specified by the amplitudes of its basis states $\ket{0}$ and $\ket{1}$ and written as $\ket{\psi}=\alpha_0 \ket{0} + \alpha_1 \ket{1}$ with $\alpha_0, \alpha_1 \in \mathbb{C}$ and $|\alpha_0|^2 + |\alpha_1|^2 = 1$.
Unfortunately, the respective $\alpha$-values cannot be directly observed.
Instead, \emph{measuring} a qubit causes its state to collapse to one of its basis states with the probability of its squared amplitude $|\alpha_i|^2$.

\begin{example}
    A qubit in an equal superposition state---both basis states occur with equal probability---can be written as $\ket{\psi}=\frac{1}{\sqrt{2}} \ket{0} + \frac{1}{\sqrt{2}} \ket{1}$.
    Measuring the state of this qubit would destroy the superposition and collapse it to either $\ket{0}$ or $\ket{1}$, both with a probability of $|\frac{1}{\sqrt{2}}|^2 = 0.5$.
\end{example}

These qubits can now be used for computation. To this end, quantum operations in the form of quantum \emph{gates} are applied that change the underlying state of the qubits.
There exists a variety of different \emph{single}- and \emph{multi-qubit} quantum gates that act on single or multiple qubits, respectively.
To realize a quantum application, end users need to generate a quantum \emph{circuit} that is comprised of multiple quantum gates acting on qubits and, based on a quantum algorithm, solves the application problem.
This quantum circuit's measurement outcome refers to the problem's solution after a possible postprocessing.

However, not only the creation of such quantum circuits is challenging, its actual \emph{execution} is as well.
Generally, there are two options: Simulators and actual quantum devices.
Simulators are inherently limited by the number of qubits they can handle, while current quantum devices suffer from execution errors that become more pronounced with increasing circuit sizes---potentially leading to unusable results.
This becomes particularly problematic for industrial-sized problems.
Thus, quantum \emph{error correction} becomes essential to compensate for these errors and to ensure the executability of large quantum circuits in a \emph{fault-tolerant} fashion.

\begin{figure*}[t]
    \centering
    \begin{subfigure}[b]{0.45\textwidth}
        \centering
        \includegraphics[width=\textwidth]{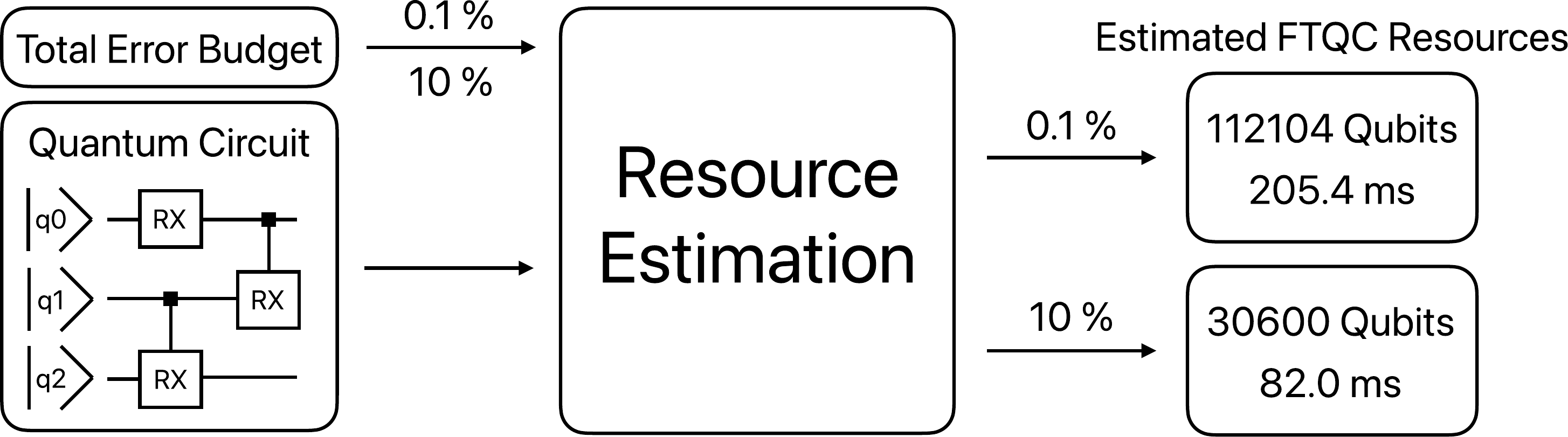}
        \caption{Resource Requirements with uniformly distributed error budgets.}
        \label{fig:default_error_budgets}
    \end{subfigure}
    \hfill
    \begin{subfigure}[b]{0.45\textwidth}
        \centering
        \includegraphics[width=\textwidth]{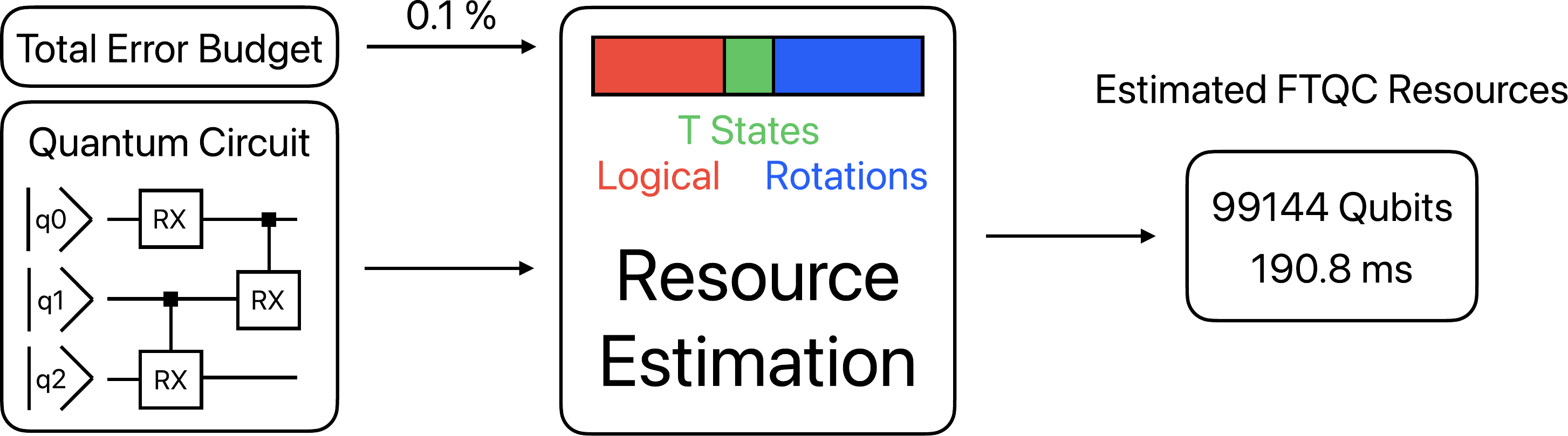}
        \caption{Resource Requirements with optimized error budgets distributions.}
        \label{fig:optimized_error_budgets}
    \end{subfigure}
    \caption{FTQC resource requirements with different error budget distributions.}
    \label{fig:FTQC_Configurations}
    \vspace{-3mm}
\end{figure*}

\subsection{Fault-Tolerant Quantum Computing}
\label{subsec:Fault-Tolerant Quantum Computing}

Similar to classical error correction (such as the repetition code~\cite{clark2013error}), quantum error correction also aims at using redundant information to correct errors. That is, one \emph{logical} qubit of a quantum circuit is actually represented by multiple \emph{physical} qubits of a quantum device. 

How exactly the logical qubit is represented by the physical qubits is determined by the chosen quantum error correction \emph{code} since qubits cannot just be copied due to the \emph{no-cloning} theorem~\cite{wootters1982single}. 
On top of that, these codes also determine how to detect and correct errors since these steps are strongly interleaved and cannot be considered independently. 
Prominent representatives of error correction codes are the surface code~\cite{fowler2012surface, litinski2019game, PhysRevLett.129.030501}, or the color code~\cite{landahl2011faulttolerantquantumcomputingcolor, PhysRevA.83.042310, Reichardt_2021, Berent2024decodingquantum}.
All codes have in common that their detection and correction mechanisms induce a significant overhead in terms of necessary physical qubits and operations (which increases the execution time).

Similar to the qubits, each logical quantum gate is represented by a code-specific sequence of physical gates, ensuring a fault-tolerant execution.
According to the \mbox{Eastin-Knill-Theorem}~\cite{PhysRevLett.102.110502}, a gate set necessary to perform any arbitrary computation---a \emph{universal} gate set---always includes both efficiently and costly implementable gates in the frame of fault-tolerance.
A popular choice of a universal gate set is the combination of Clifford gates (like H, S, and CNOT) and T gates, whereby the latter is implemented with a special quantum state, the T \emph{state}. 
It is produced in separate parts of the circuit that require substantial amounts of additional qubits and operations to implement within, e.g., the surface code.

Generally, a fault-tolerant quantum circuit consists of different parts that contribute to the overall functionality of the circuit.
One established partitioning scheme separates the circuit into the following three parts:

\begin{itemize} 
    \item \textit{Implementation of logical qubits:} This part describes the processes of representing the logical qubits and gates by physical qubits and operations.
    \item \textit{Production of T states:} As reviewed above, the T gate is the hardest to implement for which a T state is utilized that is produced in resource-intensive separate parts of the circuit.
    \item \textit{Approximation of rotation gates:} Rotation gates are single-qubit gates that rotate a qubit's state around an abstract axis and are commonly used in many quantum circuits. Since they are not part of the Clifford + T gate set, they must be approximated by a sequence of Clifford and T gates. Here, more T gates are required for a more precise approximation.
\end{itemize}

Despite the error correction techniques, errors can still occur with certain probabilities.
Naturally, these errors can happen in any of the circuit's parts, and their total sum represents the overall error of the circuit.
When reducing the probability at which errors occur, more overhead is caused in terms of resources and vice versa.
The maximum tolerated probability at which errors may occur in the result is application-specific and called \emph{error budget}.
This error budget is determined by the end user and incorporated into the design of the error-corrected quantum circuit.

\subsection{Resource Estimation}
\label{subsec:Resource Estimation}

Due to the complexity of quantum error correction and the currently limited hardware as well as software capabilities, end users cannot yet generate, simulate, and/or execute \mbox{fault-tolerant} quantum circuits for larger, i.e., practically relevant applications.
Fortunately, a complementary approach has been proposed: Resource Estimation.
Instead of creating and executing a fault-tolerant version of a given, \mbox{non-fault-tolerant} quantum circuit, it \emph{estimates} what resources in terms of the number of physical qubits and execution time would be necessary to do so.
Although this does not provide an actual result to the application problem, it allows end users to evaluate and optimize the costs of the corresponding quantum circuit (based on the estimated costs) as well as to assess whether a corresponding solution is feasible in the (near) future~\cite{10821277}.

In the past years, many powerful resource estimation tools have been proposed, such as, e.g., Microsoft's Azure Quantum Resource Estimator~\cite{van2023using, beverland2022assessingrequirementsscalepractical}, Google's Qualtran~\cite{harrigan2024expressinganalyzingquantumalgorithms}, and MIT Lincoln Lab's pyLIQTR~\cite{rroodll_2024_10913397}.
In this work and without loss of generality, we utilize the Microsoft Azure Quantum Resource Estimator.
The tool calculates the logical qubit overhead introduced through an assumed computation scheme (PSSPC~\cite{beverland2022assessingrequirementsscalepractical}) based on the logical gate counts in the input circuit.
Together with customizable aspects like quantum error correction assumptions (for all of which we assume the default settings), this constitutes the input for the tool.
Alongside this, the estimation allows for specifying an error budget (reviewed in \autoref{subsec:Fault-Tolerant Quantum Computing}) and its distribution.
Analogously to the circuit partitioning, the resource estimator divides the error budget into the logical, T states, and rotations error budget.
The sum of these individual error budgets refers to the total error budget.
If not specified differently, the resource estimator assumes a uniform distribution of the total error budget among the three error budgets.

\begin{figure*}[t]
    \centering
    \begin{subfigure}[b]{0.289\textwidth}
        \centering
        \includegraphics[width=\textwidth]{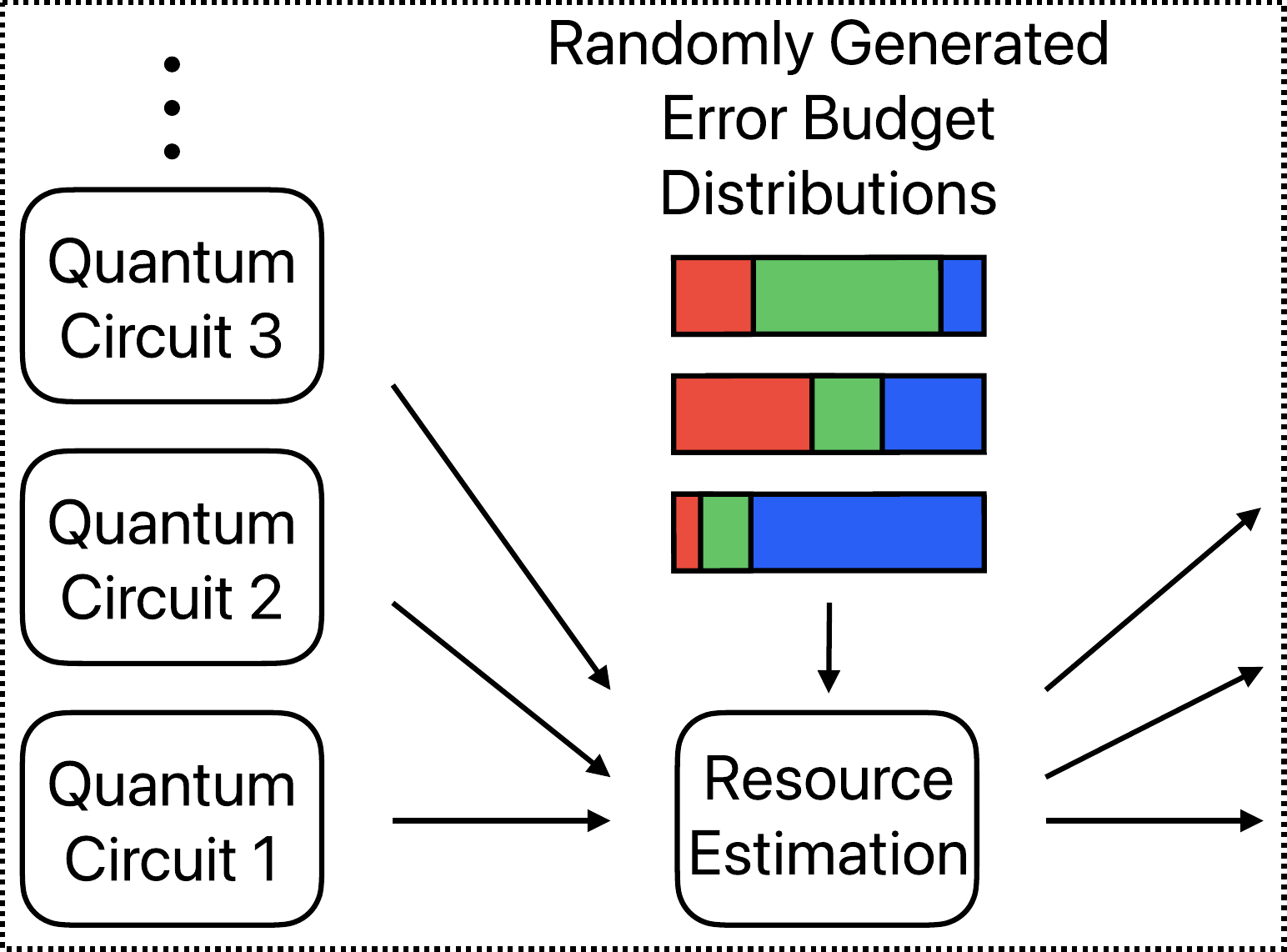}
        \caption{Sampling.}
        \label{fig:error_budget_methodology_sampling}
    \end{subfigure}
    \hfill
    \begin{subfigure}[b]{0.257\textwidth}
        \centering
        \includegraphics[width=\textwidth]{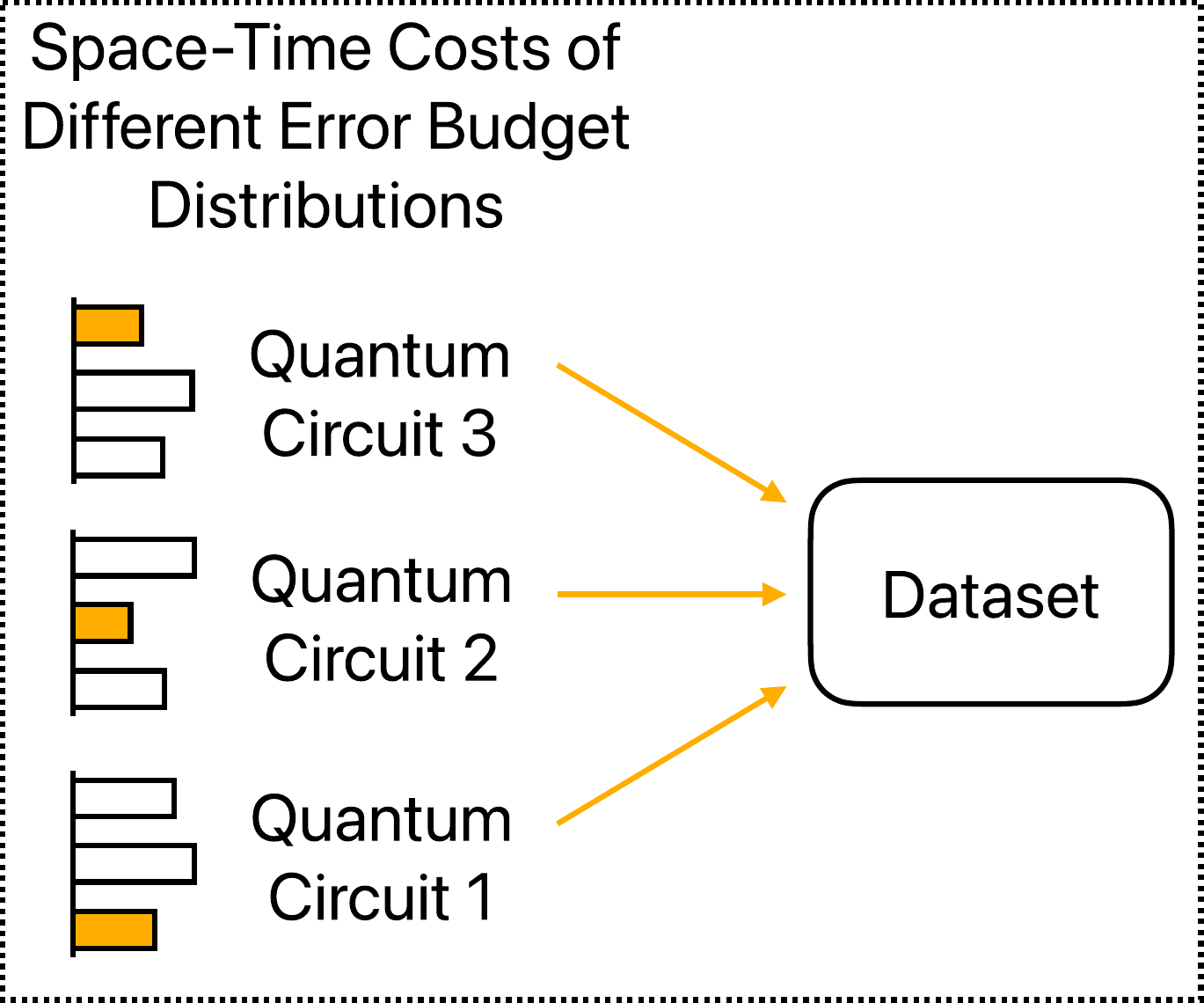}
        \caption{Data Accumulation.}
        \label{fig:error_budget_methodology_accumulation}
    \end{subfigure}
    \hfill
    \begin{subfigure}[b]{0.424\textwidth}
        \centering
        \includegraphics[width=\textwidth]{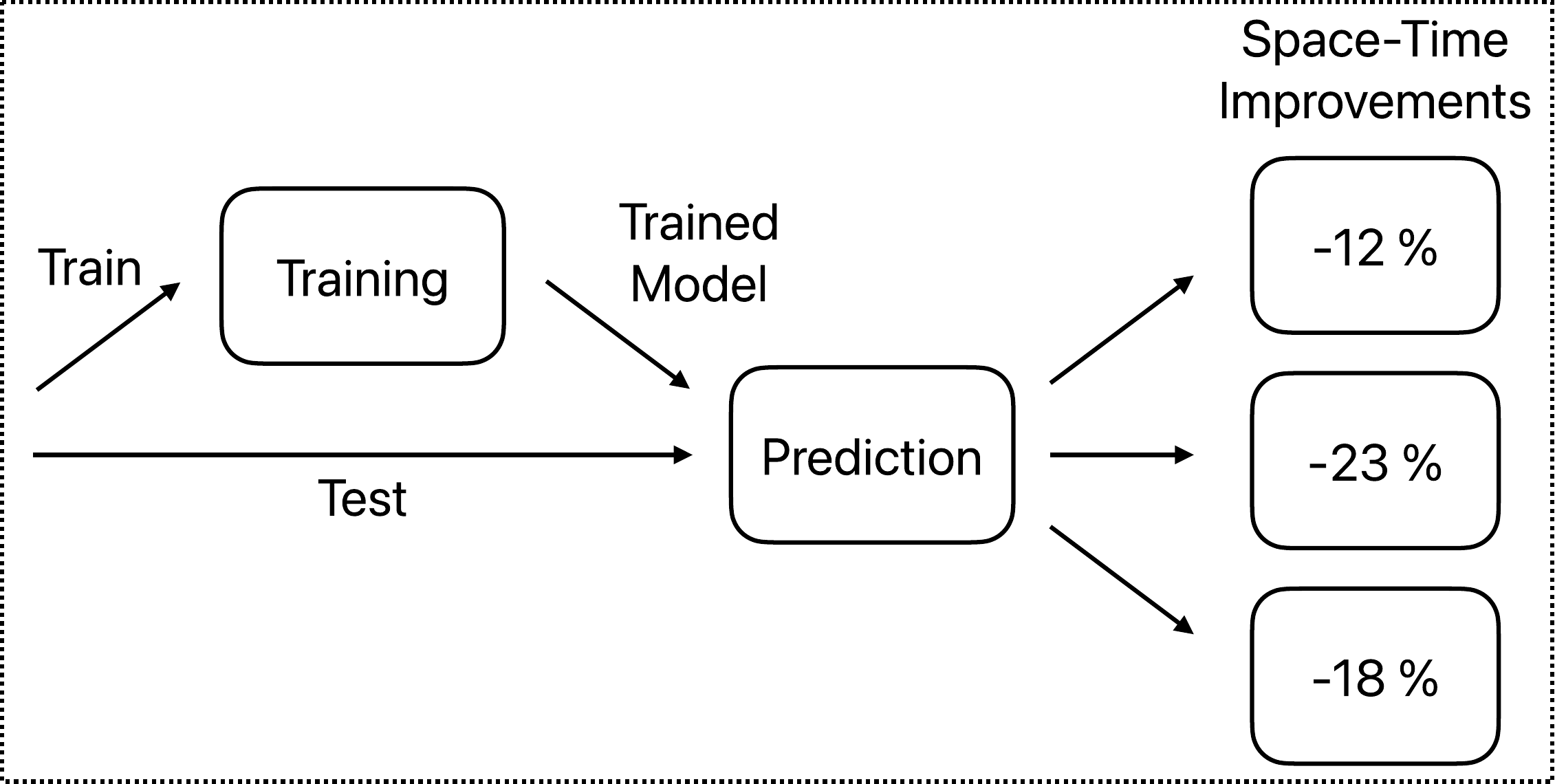}
        \caption{Training and Evaluation.}
        \label{fig:error_budget_methodology_training_eval}
    \end{subfigure}
    \caption{Proposed method.}
    \label{fig:methodology}
    \vspace{-4.5mm}
\end{figure*}

\section{Motivation}
\label{sec:Motivation}

When utilizing quantum computing, end users need to formulate their problems in a format suitable for quantum computing (i.e., a quantum circuit). 
Afterward, these quantum circuits are passed through a compilation stack that automatically transforms them into circuits that can be executed on corresponding devices~\cite{kusyk2021survey, quetschlich2025mqtpredictor, wille2023mqt}. 
Since these devices are \mbox{subject to noise}, errors are usually induced when executing a quantum circuit. 
Accordingly, quantum error correction techniques such as reviewed in \autoref{subsec:Fault-Tolerant Quantum Computing} are used to compensate for \mbox{these errors}.

But these quantum error corrections techniques induce dramatic overheads in the physical resource \mbox{requirements---often} leading to an increase in, e.g., the number of \mbox{qubits by orders} of magnitudes. 
Consequently, even small quantum circuits, initially relying on a few qubits, suddenly require millions of qubits to execute them in a fault-tolerant fashion. 
\mbox{Obviously}, this needs to be reduced substantially when aiming for \mbox{fault-tolerant} quantum computing \mbox{solutions for practical problems}.

In addition to further improvements in hardware and software, application-based approaches can contribute to reaching this goal. 
Here, an application's maximum tolerated error, the \emph{error budget}, can be a particular lever. 
In fact, many applications for quantum computing do tolerate a certain degree of errors during execution as, for example, Shor's algorithm~\cite{monz2016realization} or \mbox{noise-robust} ground state energy estimates~\cite{Vallury2023noiserobustground}. 
Hence, end users need to provide this error budget and, then, the compilation stack can accordingly take that into account. 
Then, applications with a larger error budget can be realized with fewer qubits and fewer additional operations. 
To explore these effects, resource estimation, as reviewed in \autoref{subsec:Resource Estimation} and illustrated in the following example, can be utilized.

\begin{example}
\label{ex:error budget}
    Consider the initial quantum circuit shown in \autoref{fig:FTQC_Configurations}, which consists of three qubits, one RX, and two controlled RX gates.
    When executed fault-tolerantly with an error budget of $0.1\%$, it is estimated to require $112104$ physical qubits and an execution time of $205.2$ ms.
    In contrast, when an error budget of $10 \%$ is considered, the estimated resources reduce to $30600$ qubits and $82.0$ ms.
\end{example}

However, the error budget used thus far just refers to the overall errors and assumes a \emph{uniform} distribution of errors. 
But, the circuit is separated into different parts (with a common partitioning of the circuit into the implementation of logical qubits, production of T states, and approximation of rotation gates as reviewed in \autoref{subsec:Fault-Tolerant Quantum Computing}).
Each part may handle errors in different fashions and with different overheads---which all also influence each other.
This provides a huge degree of freedom that can be used to reduce the expected costs. 

More precisely, it is possible to accept more errors in one particular part of the circuit (making the corresponding part of the quantum circuit less costly) \mbox{as long as this is} compensated by more error correction towards another part of the circuit (making this part of the quantum circuit more costly).
\mbox{\emph{Trading off} corresponding} improvements and deteriorations of costs by changing the distribution of errors (instead of assuming a uniform distribution) potentially allows \mbox{for a} much more resource-efficient construction of the overall circuit. 
As long as the total error budget provided by the end user remains fixed, the desired result (namely, a fault-tolerant circuit realizing the given application) is reached, but with significantly fewer costs.

\begin{example}
    In \autoref{ex:error budget} and \autoref{fig:default_error_budgets}, a uniform distribution of error budgets has been assumed, leading to the estimated required resources.
    When changing this distribution while keeping their sum---and, hence, the total error budget---fixed, the estimated resource requirements change. 
    \autoref{fig:optimized_error_budgets} illustrates this for the case of increasing the logical and rotations error budgets and compensating for this by reducing the \mbox{T states} error budget (indicated by the red, blue, and green bars, respectively). 
    Compared to the uniform distribution, the required physical qubits and execution time decrease by $\sim 12\%$ and $\sim 7\%$, respectively.
\end{example}

As shown, it is possible to find distributions that still meet the overall error budget but, in an attempt to reduce the estimated costs, utilize the degree of freedom in how this budget is reached. 
This (in combination with existing resource estimation methods) allows end users to better approximate the utility of their quantum circuit applications. 
Unfortunately, finding a proper error distribution is not an easy task: Since the proportions of the three individual error budgets are defined by continuous values, infinitely many different distributions are possible---making it infeasible to evaluate by end users. 
Motivated by this, the remainder of this work proposes a method that not only provides general guidelines on how to optimize an error budget distribution in general but also \emph{automatically} determines optimized error budget distributions tailored for an arbitrary quantum circuit.

\begin{figure*}[t]
    \centering
    \begin{subfigure}[b]{0.3\textwidth}
        \centering
        \includegraphics[width=\textwidth]{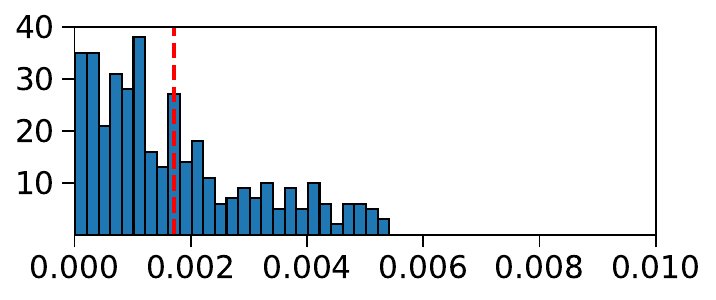}
        \caption{Logical error budget.}
        \label{fig:logical error budget}
    \end{subfigure}
    \hfill
    \begin{subfigure}[b]{0.3\textwidth}
        \centering
        \includegraphics[width=\textwidth]{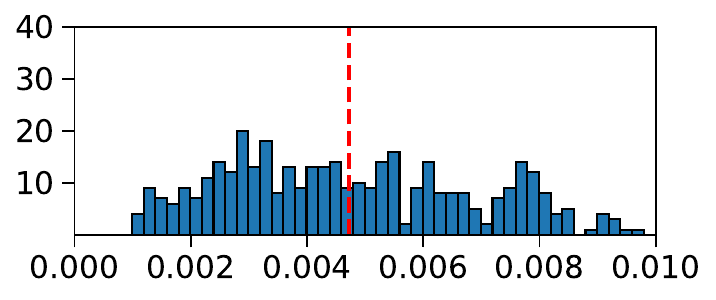}
        \caption{T states error budget.}
        \label{fig:t states error budget}
    \end{subfigure}
    \hfill
    \begin{subfigure}[b]{0.3\textwidth}
        \centering
        \includegraphics[width=\textwidth]{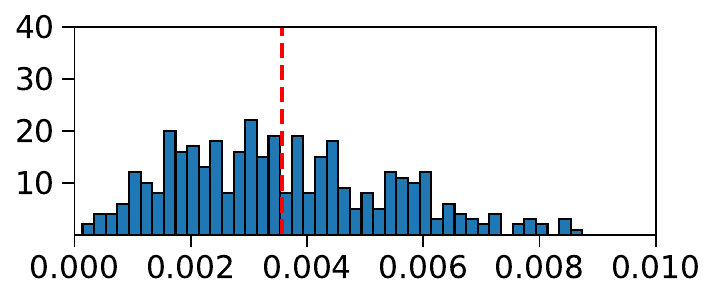}
        \caption{Rotations error budget.}
        \label{fig:rotations error budget}
    \end{subfigure}
    \caption{Frequencies of the three error budgets in the test set for a total budget of 1 \%.}
    \label{fig:Error-Budget-Distributions}
    \vspace{-5mm}
\end{figure*}

\section{Optimizing the Error Budget Distribution}
\label{sec:Optimizing the Error Budget Distribution}

This section describes the proposed approach to automatically optimize the error budget distributions for \mbox{resource-efficient} designs of \emph{fault-tolerant} quantum circuits.
To this end, the approach utilizes a supervised machine learning model that predicts such distribution based on a quantum circuit.

For this, a labeled training dataset is necessary that can be used to train a respective machine learning model.
The core of such an approach is obviously to ensure the model performs well on unseen arbitrary data---or arbitrary quantum circuits in this case.
Therefore, carefully selecting a representative training dataset that includes as much information as possible for the model to learn is necessary.

Hence, the proposed approach (sketched in \autoref{fig:methodology}) uses various different representative quantum circuits, for which different random error budget distributions are sampled and the respective resources estimated (as sketched in \autoref{fig:error_budget_methodology_sampling}).
Then, the best error budget distribution with regard to the estimated resources is chosen for each quantum circuit to accumulate a dataset consisting of quantum circuits and corresponding optimized error budget distributions (as sketched in \autoref{fig:error_budget_methodology_accumulation}).
The resulting error budget distributions in this dataset are investigated and used to indicate general guidelines that are derived based on explicit insights into what distributions led to the best results.
Furthermore, the machine learning model is trained to map the quantum circuits on the optimized error budget distributions with the dataset before evaluating it with unseen test data (as sketched in \autoref{fig:error_budget_methodology_training_eval}).
This enables tailoring an optimized distribution for an arbitrary quantum circuit based on the model's predictions.
The three steps of sampling, accumulating data, and training are described in more detail next.

\subsection{Sampling}
\label{subsec:Sampling}

Since the basis of this approach is to generate a machine learning model, creating the dataset for its training is decisive.
Considering that the model is supposed to learn the relationship between quantum circuits and corresponding resource-efficient error budget distributions, a set of different representative quantum circuits is required for this purpose.
These can be provided by the end user or taken from benchmarking databases.
As sketched in \autoref{fig:error_budget_methodology_sampling}, numerous random distributions of error budgets are sampled, and the according resources are estimated for each quantum circuit.
To comply with the condition of their sum equaling the given total error budget, this is done as follows:

\begin{equation*}
\begin{aligned}
    x_i &\sim \text{Uniform}(0,1),\; i \in \{L,T,R\} \\
    x_i &\leftarrow \frac{x_i}{\sum x_i} \cdot \text{Total Error Budget}.
\end{aligned}
\end{equation*}
Here, $x_i,\; i \in \{L,T,R\}$ refer to the logical, T states, and rotations error budget used for the subsequent step.
This sampling step results in a set of many randomly generated error budget distributions together with the according resource estimates for each quantum circuit.

\subsection{Data Accumulation}
\label{subsec:Data Accumulation}
Based on these generated distributions, a metric is needed to define what constitutes the best solution.
To determine what best actually means is a degree of freedom. 
An obvious choice is the lowest product of the estimated number of physical qubits (space) and execution time, hence the lowest estimated \mbox{\emph{space-time}} costs.
However, end users might primarily be interested in the qubit count or required execution time.
Because the proposed approach aims to train a supervised machine learning model, accumulating a labeled dataset is required.
A numerical representation format for characterizing quantum circuits is essential to enable the model to learn the relationship between a quantum circuit and the corresponding optimized distribution.
The logical counts provided by Microsoft's resource estimator, reviewed in \autoref{subsec:Resource Estimation}, are well-suited for this task.
Thus, after every quantum circuit has been iterated, the accumulated dataset consists of the logical counts of each quantum circuit and the according error budget distribution that leads to optimized \mbox{space-time} costs (or whatever other cost function is chosen).
This dataset can then be used to analyze the resulting optimized error budget distributions and train a machine learning model to predict those.

\subsection{Training}
\label{subsec:Training and Evaluation}

After accumulating the dataset, a machine learning model is trained to learn the relationship between quantum circuits and the corresponding optimized error budget distributions.
After the training, the model can be used to predict preferable distributions for arbitrary quantum circuits, enabling end users to feed their quantum circuits into the model and obtain optimized error budget distributions.
By varying the total error budget as well as the metric for selecting the best distribution introduced above, different machine learning models can be trained to cover multiple different use cases.
To evaluate the effectiveness of the proposed approach, the machine learning model needs to be tested on so far unseen but labeled data.
To this end, the accumulated dataset is split into a train and test set.
The latter is used to provide a comprehensive evaluation of the proposed approach in the following section. 

\begin{figure*}[t]
    \centering
    \begin{subfigure}[b]{0.28\textwidth}
        \centering
        \includegraphics[width=\textwidth]{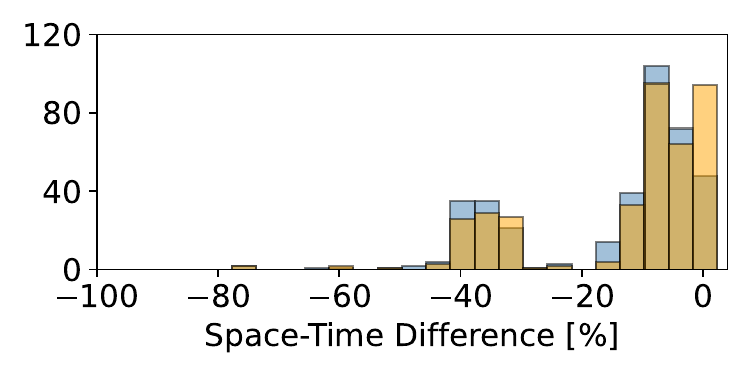}
        \caption{Total error budget of 0.1 \%.}
        \label{fig:space-time-improvements-0.1}
    \end{subfigure}
    \hfill
    \begin{subfigure}[b]{0.28\textwidth}
        \centering
        \includegraphics[width=\textwidth]{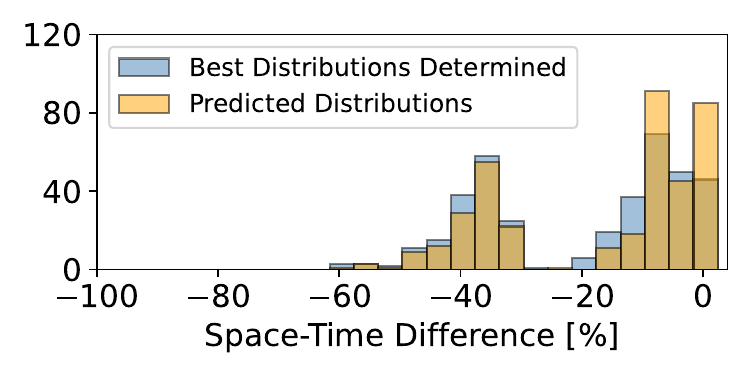}
        \caption{Total error budget of 1 \%.}
        \label{fig:space-time-improvements-1}
    \end{subfigure}
    \hfill
    \begin{subfigure}[b]{0.28\textwidth}
        \centering
        \includegraphics[width=\textwidth]{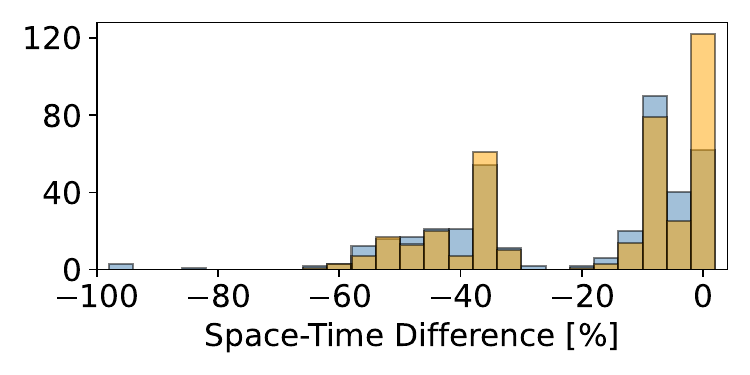}
        \caption{Total error budget of 10 \%.}
        \label{fig:space-time-improvements-10}
    \end{subfigure}
    \caption{Improvements in the estimated space-time costs by the proposed approach.}
    \label{fig:Improvements}
    \vspace{-5mm}
\end{figure*}

\section{Evaluation}
\label{sec:Evaluation}

To demonstrate its effectiveness, a comprehensive evaluation of the proposed approach covering various use cases in the form of different representative quantum circuits and total error budgets was conducted.
This provides general \mbox{guidelines---obtained} from the accumulated dataset---on how to distribute the error budget in a resource-efficient way.
Moreover, the reductions in the estimated \mbox{space-time} costs achieved by individually tailoring the error budget distribution with the machine learning model are described.
In this section, all these obtained results are summarized.
Before, a brief description of the corresponding applied setup is described.
Note that, in addition, the implementation is available on GitHub (\url{https://github.com/cda-tum/mqt-problemsolver}) as part of the \emph{Munich Quantum Toolkit}~\cite{willeMQTHandbookSummary2024} to enable end users to implement their own setups and evaluations.

\subsection{Setup}
\label{subsec:Setup}

In total, 1530 different quantum circuits from the \mbox{MQT Bench library~\cite{quetschlich2023mqtbench}} (representing different quantum algorithms in various sizes in both the number of qubits and gates) were considered.
For each quantum circuit, 1000 error budget distributions were randomly sampled.
To ensure the model's predicted error budget distributions meet the desired total error budget, they were normalized before the resource estimation.
When evaluating the machine learning model, the error budget distribution with the lowest estimated \mbox{space-time} cost was chosen from the machine learning model's predictions and a uniform distribution.

To conduct the resource estimation, Microsoft's Azure Quantum Resource Estimator~\cite{van2023using, beverland2022assessingrequirementsscalepractical} (version 1.9.0) was used with default parameters apart from the error budgets.
As a baseline, the estimates were compared to a uniform distribution by only restricting the total error budget in the resource estimator.
Once every quantum circuit had been iterated, the dataset was split into a train (1147 samples or $75 \%$) and a test set (383 samples or $25 \%$) after a random shuffling. 
As a machine learning model, a Random Forest solution was implemented with scikit-learn~\cite{scikit-learn} (version 1.5.2).
This procedure was conducted for the total error budgets $0.1 \%$, $1 \%$, and $10 \%$, while the data for each total error budget was produced with its own trained machine learning model.

\subsection{Obtained Guidelines on Distributing the Error Budget}
\label{subsec:Detailed Insights}
The approach proposed above allows one to get insights into what, for a given total error budget, an optimized distribution to the respective parts (logical, T states, and rotations budget) should look like.
To this end, the optimized error budget distributions determined for the different quantum circuits in the accumulation step were investigated.
The histograms provided in \autoref{fig:Error-Budget-Distributions} illustrate how frequently a certain value of the logical, T states, and rotations error budget was picked as the most \mbox{space-time} efficient in the test set.
The dashed red lines refer to the mean values of each budget, with their sum equaling the total error budget.
For this evaluation, we focus on the $1 \%$ total error budget case while the other two cases behave similarly.

\autoref{fig:logical error budget} shows this for the case of the logical error budget.
Since most of the blue bars are concentrated near zero, small values appeared significantly more often in the dataset than large values, which is reflected by the mean value of around 0.0017.
A logical error budget above half of the total error budget appeared almost never in the dataset. 
Conclusively, smaller error budgets for this part of the circuit tend to be preferable for \mbox{space-time} efficient designs of fault-tolerant quantum circuits.

The corresponding histograms for the T states and rotations error budgets are provided in \autoref{fig:t states error budget} and \autoref{fig:rotations error budget}, respectively.
Both error budgets showed a widely spread occurrence over different proportions of the total budget in the test set, with a slight shift to higher values for the T states budget compared to the rotations budget. 
Their mean values are around 0.0047 and 0.0036 for the T states and rotations budget, respectively.
Based on these results, the following guidelines on how to distribute the error budget can be drawn:

\begin{itemize}
    \item It is, on average, more resource-efficient to reduce the error budget in implementing the logical qubits and allowing for more errors in the production of T states compared to a uniform distribution.
    \item Since the approximation of rotation gates includes the utilization of T states, the rotations error budget behave similarly to the T states budget.
    \item Allowing for more errors in this approximation and compensating for this with lower errors in implementing the logical qubits is, on average, more resource-efficient than a uniform distribution.
\end{itemize}
Despite these conclusions indicating rough guidelines for the distribution of error budgets in a fault-tolerant quantum circuit, it is essential to consider each circuit individually for this task.
Especially the optimized T states and rotations error budget differ significantly across the samples.
This can be explained by the considerable differences in gate counts, especially T and rotation gates, across the samples.
Hence, an individually tailored error budget distribution for each quantum circuit is essential for designing a resource-efficient fault-tolerant quantum circuit.
This is achieved by the trained machine learning model, which is used to predict optimized error budget distributions for arbitrary quantum circuits as proposed above.
The resulting resource estimates by these predicted distributions compared to a uniform distribution are described next.

\subsection{Comparison to Uniform Distribution}
\label{subsec:Overall Results}

Next, the obtained error budget distributions of the proposed approach are compared to the usually assumed uniform distribution for the quantum circuits of the test set.
The results are summarized in \autoref{fig:Improvements} in the form of histograms for three different total error budgets.
The plots illustrate the best distributions determined (blue) for quantum circuits of the test set during the accumulation step described above and the machine learning model's predictions when inputting the respective quantum circuits (yellow).
Here, the x-axis refers to the change in the estimated \mbox{space-time} costs compared to a uniform error budget distribution in percent, while the y-axis refers to the frequency with which certain values appeared in the data.
In the following, the rounded results are always described in the order of total error budgets of $0.1 \%$, $1 \%$, and $10 \%$ as illustrated in \autoref{fig:space-time-improvements-0.1} to \autoref{fig:space-time-improvements-10}, respectively.
The estimated \mbox{space-time} costs could be reduced for over $77 \%$, $79 \%$, and $70 \%$ of the considered samples with an average reduction of $12.8 \%$, $16.3 \%$, and $17.7 \%$.
Samples with the maximum reductions achieved an estimated difference in the \mbox{space-time} costs of $77.7 \%$, $61.6 \%$, and $62.3 \%$.

On average, across all total error budgets and quantum circuits, a reduction of $15.6 \%$ could be achieved with predictions of the machine learning model compared to a uniform distribution (which is usually assumed), including samples without improvement.
Over $75 \%$ of the quantum circuits yielded improvements in the estimated \mbox{space-time} costs with this approach.
Overall, the results of the predictions of the machine learning models resembled the best distributions determined by the proposed approach.
The similarities in the results across the different total error budgets underline the customizability of the approach to fit a broad variety of use cases.

\section{Conclusion}
\label{sec:Conclusion}

Executing quantum circuits on real quantum devices results in significant noise for practically relevant problems due to quantum errors. 
Therefore, quantum error correction becomes necessary to handle these errors but induces substantial overheads in physical qubits and execution time. 
To reduce these overheads, the maximum tolerated error for a quantum application, the error budget, can help. 
For this, the error budget can be distributed such that circuit parts that can compensate for errors more efficiently than others contribute more to the total error budget.
However, finding such distributions is challenging.
In this work, an approach has been proposed that addresses this challenge by accumulating a dataset used to draw general guidelines for distributing the error budget.
Furthermore, a machine learning model has been trained to predict optimized distributions for arbitrary circuits.
The effectiveness of the proposed approach has been confirmed in an experimental evaluation in the form of a comparison to the usually assumed uniform distribution.
Here, the estimated \mbox{space-time} costs for over $75 \%$ of the considered samples have been reduced with an average reduction of $15.6 \%$ (including samples without improvement) and a maximum reduction of $77.7 \%$.
Future work includes extending the approach to more use cases and further investigating the relationship between quantum circuit characteristics and resource-efficient error budget distributions, both empirically and theoretically.

\section*{Acknowledgments}

This work received funding from the European Research Council (ERC) under the European Union’s Horizon 2020 research and innovation program (grant agreement No. 101001318), was part of the Munich Quantum Valley, which is supported by the Bavarian state government with funds from the Hightech Agenda Bayern Plus, and has been supported by the BMK, BMDW, the State of Upper Austria in the frame of the COMET program, and the QuantumReady project within Quantum Austria (managed by the FFG).

\printbibliography

\end{document}